\documentclass[12pt]{article}
\usepackage{hyperref}
\usepackage{amsmath}
\usepackage{mathrsfs}
\usepackage{graphicx}
\usepackage{hyperref}
\usepackage{amssymb}
\usepackage{mathrsfs}
\usepackage{amsthm}
\usepackage{graphicx}
\usepackage{xcolor}
\usepackage{hyperref}
\usepackage{amsmath}
\usepackage{graphicx}
\usepackage{mathrsfs,stmaryrd,mathtools}
\usepackage{multicol}

\begin{document}

 \title{\vskip-2cm\large \bf Probing the Black Hole Singularity with a Quantum String}

 \author{\normalsize {\sc Harpreet Singh\footnote{\tt harpreetsingh@iitg.ac.in}\ \ and  Malay K. Nandy\footnote{{\tt mknandy@iitg.ac.in} (Corresponding Author)}}\\{ \normalsize \em Department of Physics, Indian Institute of Technology Guwahati}\\ {\normalsize \em Guwahati 781 039, India}}

 \date{\normalsize (7 October 2026)\footnote{Revised version of Report dated 8 August 2026.}}

\maketitle
 
\begin{abstract}

In this paper, we investigate the quantum behaviour near the classical singularity of a black hole. We describe the black-hole interior by the Kantowski-Sachs metric, with a string propagating in this region.  This approach is motivated by the fact that, near the classical singularity, spacetime curvature reaches Planck or string scales, necessitating a description in terms of string-theoretic degrees of freedom.  The string dynamics is formulated via the Polyakov action, and the combined system is canonically quantized to derive the corresponding Wheeler-DeWitt equation.  Although the equation is separable in the radial and azimuthal string sectors, the remaining geometric and string degrees of freedom are nontrivially coupled.  To address this, we employ a perturbative approach and obtain the wave function up to first order.  At the zeroth-order, the angular part is described by the general Heun function, while the geometric part is expressed in terms of the Bessel function.  The first-order wave function is governed by a source term containing the zeroth-order solution, and we obtain its explicit expression in terms of the complementary function and the particular integral.  We confirm  regularity of the wave function at both zeroth and first order, at the singularity as well as at the horizon. Regularity at the horizon imposes a discretization condition on one eigenvalue, while another eigenvalue is quantized due to the azimuthal periodicity of the wave function.  Importantly, both zeroth- and first-order solutions satisfy the DeWitt boundary condition, implying  resolution of the black hole singularity within this quantum framework.\\

\noindent Keywords: Quantum gravity; Black hole singularity resolution; Kantowski-Sachs mini\-superspace; Polyakov string action; Wheeler-DeWitt equation; DeWitt boundary condition

\end{abstract}

\pagebreak

\tableofcontents

\section{Introduction}

One of the most profound indications of the incompleteness of classical general relativity is the occurrence of spacetime singularities. The singularity theorems of Penrose and Hawking demonstrate that, under broad physical assumptions, gravitational collapse inevitably leads to geodesic incompleteness and the formation of singular regions where classical descriptions cease to be valid \cite{Penrose1965,Hawking1970,HawkingEllis1973,Wald1984,MisnerThorneWheeler1973,Geroch1968,Chandrasekhar1983,OppenheimerSnyder1939}. In particular, the Schwarzschild black hole solution contains a spacelike singularity in its interior where curvature invariants diverge and the predictive power of classical gravitational theory breaks down. The existence of such singularities is generally regarded as evidence that a quantum theory of gravity is required to describe the fundamental structure of spacetime in the deep regime of extreme curvature.

Black holes therefore provide an important arena for investigating quantum gravitational effects. Since the interior of a black hole is causally disconnected from distant observers and inevitably evolves toward the classical singularity, it offers a natural setting in which the limitations of classical gravity become manifest. Understanding the fate of the interior singularity has consequently become a central objective of several approaches to quantum gravity. Over the years, a variety of mechanisms for singularity resolution have been proposed, including those arising in canonical quantum gravity, loop quantum gravity, string theory, and related frameworks \cite{Wheeler1963LHL,DeWitt1967,Ashtekar2005,Bojowald2005,Polchinski_1998,Zwiebach2009,Rovelli2008,GambiniPullin2013,AshtekarSingh2011,Modesto2006,LinYeomStojkovic2025}.

Within canonical quantum gravity, the dynamics of spacetime is governed by the Wheeler-DeWitt equation, obtained by imposing the Hamiltonian constraint at the quantum level \cite{Wheeler1963LHL,DeWitt1967,Kiefer2012,Wheeler1968,Isham1991,Dirac1958}. In this approach, the quantum state of a gravitational system is represented by a wave functional defined on superspace. Because it becomes intractable to consider the full superspace of general relativity, investigations are often carried out within minisuperspace models obtained by imposing suitable symmetry reductions. Such models retain essential gravitational degrees of freedom while rendering the quantum theory tractable. The Schwarzschild black hole interior geometry admits a description in terms of the Kantowski-Sachs metric \cite{Misner1972,Ryan1975,KantowskiSachs1966,Conradi1997,ChibaMatsuiMurata2026,AsafeWDW2026}, which has therefore served as a useful framework for studying quantum aspects of black hole interiors \cite{Kuchar1994,Ashtekar2005,SinghNandy2025}.

A longstanding question in canonical quantum gravity concerns the characterization of classical singularities within a quantum framework. One influential proposal was introduced by DeWitt, who suggested that physically admissible quantum states should vanish on configurations corresponding to classical singularities \cite{DeWitt1967}. According to this criterion, if the wave function approaches zero in the singular region of configuration space, the corresponding singular geometry is excluded from the physical quantum state space. The DeWitt boundary condition has subsequently been employed in numerous investigations of black hole and cosmological singularities and continues to provide a useful diagnostic for assessing singularity resolution in Wheeler-DeWitt quantization schemes \cite{Kiefer2019,Kiefer2012,Malafarina2014,BouhmadiLopez2014,CraigSingh2013,GotayDemaret1983,GielenRied2025}. While this criterion provides a useful characterization of singularity resolution from the perspective of quantum geometry, it is also important to examine whether additional degrees of freedom, beyond those of the spacetime geometry itself, can influence the quantum behaviour in the strong-curvature regime.

In this regard, it is anticipated that the physics near spacetime singularities cannot be fully captured by conventional point-particle descriptions. As the curvature approaches the Planck or string scale, extended degrees of freedom are expected to become important. String theory provides a natural framework in which such effects can be incorporated, replacing point-like probes with one-dimensional strings whose dynamics are governed by the Polyakov action \cite{Polyakov1981,Green1987,Polchinski_1998,Zwiebach2009,HorowitzSteif1991,Larsen1997,Witten1991,Callan1992,Veneziano1991,Wu2026}. Studies have suggested that string-theoretic effects can significantly modify the structure of spacetime in strong-curvature regimes and may contribute to the resolution of classical singularities \cite{Horowitz1991,Larsen1997,Kanai_2026}. In addition, quantum string-theoretic considerations \cite{Singh2025str} in the exterior classical background of an AdS$_5$-Schwarzschild black hole have led to the emergence of a trapping surface in the exterior region, analogous to the stretched horizon proposed by Susskind in the context of black hole complementarity \cite{Susskind1993a}. These results suggest that extended string degrees of freedom may provide useful perspectives on the quantum structure of black hole spacetimes, particularly in regimes where the classical geometric description becomes inadequate.

Motivated by these considerations, it is natural to investigate quantum gravitational models in which geometric and string degrees of freedom are treated simultaneously. Such a framework makes it possible to examine how an extended object propagates within a quantum black hole interior and, in particular, to explore whether its dynamics modify the behaviour of the quantum state near the classical singularity. The string degrees of freedom can thereby serve as a probe of the deep quantum regime, where semiclassical descriptions are expected to become unsuitable \cite{DEVEGA1987320,DIJGRAAF1992269,DEVEGA1988577,Horowitz1998,Susskind1993,Vega1996}. This provides a natural setting for examining the interplay between quantum geometry and extended string degrees of freedom in the vicinity of the black hole singularity.

In the present work, we therefore investigate the quantum dynamics of a string propagating in the interior of a black hole described by the Kantowski-Sachs geometry. To this end, we start from the Einstein-Hilbert action for the Kantowski-Sachs geometry together with the Polyakov action for the string sector, construct the Hamiltonian formulation of the coupled system, and perform its canonical quantization. The resulting Wheeler-DeWitt equation incorporates both geometric and string degrees of freedom, thereby providing a unified quantum description of the coupled geometry-string system. This framework allows us to examine, in particular, how the presence of an extended string degree of freedom influences the quantum behaviour of the black hole interior in the vicinity of the classical singularity.

We analyze the resulting Wheeler-DeWitt equation by separation of variables, obtaining exact solutions for the decoupled sectors of the wave function. For the remaining coupled sectors, we develop a perturbative treatment and obtain the corresponding zeroth- and first-order solutions. Their behaviour is then examined both in the vicinity of the classical singularity and near the black hole horizon. The DeWitt criterion is employed to assess singularity resolution, regularity conditions at the horizon are used to constrain the admissible quantum states.

We find that that the wave function vanishes at the classical singularity and remains well-behaved under perturbative corrections, suggesting a quantum resolution of the black hole singularity in the presence of string degrees of freedom.  These results indicate that the the coupled geometry-string scenario provides a suitable framework for quantum resolution of the black hole singularity.

The remainder of the paper is organized as follows. In Sec.~\ref{classicalBH}, we introduce the classical Hamiltonian formulation of the black hole interior described by the Kantowski-Sachs metric and find the corresponding Hamiltonian. In Sec.~\ref{classicalstr}, we obtain the Hamiltonian of a string propagating in the black hole interior through the Polyakov action. Sec.~\ref{QBHS} is devoted to the canonical quantization of the of the coupled gravity-string system, leading to the Wheeler-DeWitt equation describing the coupled quantum system comprising of the interior geometric and string degrees of freedom.  In Sec.~\ref{SWDW}, the Wheeler-DeWitt equation is analyzed using separation of variables, and a perturbative framework is developed to investigate the quantum behaviour near the classical singularity. The zeroth-order solutions are obtained in Sec.~\ref{zero}, where their properties are examined with particular emphasis on singularity resolution and horizon regularity. In Sec.~\ref{FOC}, first-order perturbative corrections to the wave function are calculated, including both the complementary-function and particular-integral contributions, and their implications for the quantum resolution of the singularity are discussed. Finally, the main results and their physical significance are summarized in Sec.~\ref{DC}.

\section{Classical dynamics of the black-hole interior geometry}
\label{classicalBH}

We consider the interior of the black hole to be described by the Kantowski-Sachs metric of the form 
\begin{equation}
 ds^2=-N^2(t) \:dt^2+u(t) \:dr^2+v(t) \: (d\theta^2+\sin^2\theta\:d\varphi^2),
 \label{metric}
\end{equation}
where $N(t)$ is the lapse function.
This metric \ref{metric} is already in the  ADM decomposed form, with elements 
$
 g_{\mu\nu}={\rm diag}\left(-N^2, h_{ij}\right)
$,
where $h_{ij}$ is the spatial metric with components
$
 h_{ij}={\rm diag}\left(u,v, v\: \sin^2\theta\right)
$.
The spatial Ricci scalar for the metric \ref{metric} turns out to be 
$
 ^{(3)}R=\frac{2}{v}.
$

The 
Einstein-Hilbert action 
$
 S_g=\frac{1}{16 \pi}\int \sqrt{-g}\: ^{(4)}R \,\:d^4 x
$
yields the Lagrangian in terms of the extrinsic curvature as 
\begin{equation}
L_g=\frac{1}{16\pi }\sqrt{-g}\:(K_{ij}K^{ij}-K^2+ {}^{(3)}R),
\label{lag0}
\end{equation}
in the units of $c=G=1$.

Components of the extrinsic curvature can be calculated as
$
 K_{ij}=-\frac{1}{2N}\frac{d h_{ij}}{dt}=-\frac{1}{2N}{\rm diag}\left(\dot u,\dot v,\dot v\: \sin^2\theta\right).
$
Thus, the trace of the extrinsic curvature is given by
$
 K=h^{ij}K_{ij}=-\frac{1}{2 N}\: \frac{\dot u}{u}-\frac{1}{ N}\: \frac{\dot v}{v}.
$
Moreover, the self contraction of the extrinsic curvature reads
$
 K_{ij}K^{ij}=\frac{1}{4 N^2}\frac{\dot u^2}{u^2}+\frac{1}{2 N^2}\frac{\dot v^2}{v^2}.
$
Substituting these expressions in equation \ref{lag0} gives the  Lagrangian 
\begin{equation}
L_g=\frac{\sqrt{N^2\: u\: v^2\:\sin^2\theta\:}}{16\pi}\left\{-\frac{1}{2 N^2}\frac{\dot v^2}{v^2}-\frac{1}{ N^2}\frac{\dot u\dot v}{uv}+ \frac{2}{v}\right\}.
\end{equation}

Canonical momenta conjugate to the metric degrees of freedom $u$ and $v$ are given by 
\begin{align}
P_u=\frac{\partial L_g}{\partial \dot{u}}=-\frac{\sqrt{N^2\: u\: v^2\:\sin^2\theta\:}}{16\pi }\left[\frac{1}{N^2}\frac{\dot v}{uv}\right]
\end{align}
and 
\begin{align}
 P_v=\frac{\partial L_g}{\partial \dot{v}}=-\frac{\sqrt{N^2\: u\: v^2\:\sin^2\theta\:}}{16\pi}\frac{1}{N^2}\left[\frac{\dot v}{v^2}+\frac{\dot u}{uv}\right].
\end{align}

Consequently, the Hamiltonian, $ H_g=  P_u \dot u+  P_v \dot v- L_g$, takes the form
 \begin{equation}
 H_g=\frac{16 \pi  N}{\sin \theta} \frac{1}{\sqrt{ u\: v^2\:}}\left[-uv  P_u  P_v+\frac{1}{2} (u P_u)^2-\frac{\sin^2\theta}{128\pi^2 }uv\right],
 \label{hg}
\end{equation}
in the gravity sector of the black hole interior.

\section{Classical dynamics of a  string inside the black hole}
\label{classicalstr}

The interior region of a black hole, particularly in the vicinity of the classical singularity, is expected to constitute a regime where the curvature of spacetime approaches the Planck or string scale and the classical description provided by general relativity ceases to be valid. In this domain, intrinsic quantum fluctuations of the gravitational field are anticipated to become significant and are expected to interact nontrivially with the quantum fluctuations of matter fields. Since a consistent treatment of such a regime requires a framework capable of incorporating both matter and gravitational degrees of freedom on an equal footing, it is natural to employ string-theoretic degrees of freedom. In string theory,  the spectrum of string excitations contains not only matter fields but also the graviton, thereby providing a unified description of matter and geometry. Moreover, the finite spatial extent of strings is expected to become particularly relevant in regions of extremely high curvature, where the point-particle approximation is likely to break down.

Consequently, a string propagating in the black hole interior may be regarded as a natural probe of the underlying quantum structure of spacetime and, at the same time, as an effective carrier of quantum-gravitational degrees of freedom. The study of a string coupled to the interior geometry therefore offers a useful framework for investigating how strong-curvature effects, quantum fluctuations of spacetime, and matter degrees of freedom influence one another in the deep quantum regime.

Motivated by these considerations, we consider a string propagating inside the black hole and interacting with the interior spacetime geometry described by the metric (\ref{metric}). As a first step, the coupled system is formulated at the classical level through the Polyakov action, following which a canonical quantization of the combined geometry-string system is performed in order to explore  quantum modifications of the classical singular behavior.

String dynamics is described by the Polyakov action
\begin{equation}
 S_s=-\frac{1}{4 \pi \alpha^{\prime}}\int_M d\tau d\sigma \sqrt{-h} h^{ab} g_{\mu\nu}\partial_a X^\mu \partial_b X^\nu, 
 \label{PolyakovS}
\end{equation}
where $X^\mu\equiv(t,r,\theta, \phi)$ is the  string worldsheet coordinate. The metric $h_{ab}$ characterizes the two-dimensional  worldsheet manifold $M$, with the indices $(a,b) \in \{\tau, \sigma\}$. The Regge slope $\alpha^{\prime}$ has its origin in the  string tension, $T=(2 \pi \alpha^{\prime})^{-1}$.

The two dimensional string worldsheet, along with having two reparametrization gauge invariances, has an additional symmetry with respect to Weyl (conformal) rescaling, leaving $\sqrt{-h}h^{ab}$ invariant. This makes it possible to choose a gauge in which $h_{ab}$ is diagonal,  with $h_{ab}={\rm diag}(-1,+1)$  \cite{Polchinski_1998,Polyakov1981, Green1987,Becker_Becker_Schwarz_2006}.

We shall study the dynamics of a  string  with the embeddings
$
 t=t(\tau),\  r=r(\tau),\ \theta=\theta(\tau), \ \phi=\phi(\tau).
$
With these embeddings,
the Polyakov Lagrangian in the background metric $g_{\mu\nu}$ \ref{metric}  turns out to be  
\begin{align}
 L_s=-\frac{1}{4 \pi \alpha^{\prime}}\left[N^2-u\dot r^2-v \dot \theta^2-v\sin^2\theta \dot\phi^2\right]
\end{align}
in the temporal gauge $\tau=t$.

The generalized canonical momenta conjugate to the string coordinates are obtained as
$
 P_r=\frac{\partial L_s}{\partial \dot r}=\frac{1}{2 \pi \alpha^{\prime}} u  \dot r,
$
$
 P_\theta=\frac{\partial L_s}{\partial \dot \theta}=\frac{1}{2 \pi \alpha^{\prime}} v  \dot \theta,
$
and
$
 P_\phi=\frac{\partial L_s}{\partial \dot \phi}=\frac{1}{2 \pi \alpha^{\prime}} v \sin^2\theta  \dot \phi.
$
Thus the 
string Hamiltonian $H_s=P_r\dot r +P_\theta \dot \theta +P_\phi \dot \phi-L_{s}$  is obtained in the  form 
\begin{align}
 H_s=\pi \alpha^{\prime}\left[\frac{ N^2}{4 \pi^2 \alpha'^2}+\frac{P_r^2}{ u}+\frac{P_\theta^2}{ v}+\frac{P_\phi^2}{v \sin^2\theta}\right].
\label{hamstring}
 \end{align}

\section{Wheeler-DeWitt Quantization of the black hole-string system}
\label{QBHS}

From the preceding considerations giving equations \ref{hg} and \ref{hamstring}, the total classical Hamiltonian of the black hole with a string in the interior is obtained as
\begin{align}\nonumber
 H=H_g+H_s=&\frac{16 \pi N}{\sin \theta} \frac{1}{\sqrt{ u\: v^2\:}}\left[-uv P_u P_v+\frac{1}{2} (uP_u)^2-\frac{\sin^2\theta}{128\pi^2 }uv\right]\\&+\pi \alpha^{\prime}\left[\frac{ N^2}{4 \pi^2 \alpha'^2}+\frac{P_r^2}{ u}+\frac{P_\theta^2}{v}+\frac{P_\phi^2}{v \sin^2\theta}\right]~,
 \label{classicalH}
\end{align}
with the Hamiltonian constraint $H=H_g+H_s=0$.

In the procedure of  canonical quantum gravity, the gravity-string system needs to be quantized so that the total Hamiltonian becomes an operator satisfying the Wheeler-DeWitt equation  
\begin{align}
\hat H \Psi=(\hat H_g+\hat H_s)\Psi=0.
\label{wdw1}
\end{align}

In the representation of background geometric variables and string coordinates, the conjugate momenta are promoted to operators:
\begin{equation}
\hat P_u=-i\hbar \frac{\partial}{\partial u}, \quad
\hat P_v=-i\hbar \frac{\partial}{\partial v}, \quad
\hat P_r=-i\hbar \frac{\partial}{\partial r}, \quad
\hat  P_\theta=-i\hbar \frac{\partial}{\partial \theta}, \quad
\mbox{and}  \quad
\hat P_\phi=-i\hbar \frac{\partial}{\partial \phi}.
\end{equation}
Making this transition to operators in equation \ref{classicalH}, we obtain the Wheeler-DeWitt equation from \ref{wdw1} in the differential form
\begin{align}\nonumber
 &\left[\frac{16 \pi N}{\sin \theta} \frac{1}{\sqrt{ u\: v^2\:}}\left(uv  \frac{\partial^2}{\partial u \partial v}-\frac{1}{2} u^2 \frac{\partial^2}{\partial u^2}-\frac{1}{2} u \frac{\partial}{\partial u}-\frac{\sin^2\theta}{128\pi^2 \hbar^2}uv\right)\right.\\&\left.- \pi \alpha^{\prime}\left(-\frac{ N^2}{4 \pi^2 \alpha'^2}+\frac{1}{ u}\frac{\partial^2}{\partial r^2}+\frac{1}{ v}\frac{\partial^2}{\partial \theta^2}+\frac{1}{ v \sin^2\theta}\frac{\partial^2}{\partial \phi^2}\right)\right]\Psi(u,v,r,\theta,\phi)=0.
 \label{wdw2}
\end{align}

This equation describes the coupled quantum system comprising of the interior geometric and string degrees of freedom.

\section{Solution of the Wheeler-DeWitt equation}
\label{SWDW}

\subsection{Separation of variables}

The Wheeler-DeWitt equation \ref{wdw2} is readily separable  in $r$ and $\phi$ coordinates of the string.
In order to separate the variable $r$, we first multiply equation \ref{wdw2} by $u$, and
write $\Psi(u,v,r,\theta,\phi)= \chi (u,v,\theta,\phi) R(r)$,  leading to  two differential equations of the forms
\begin{align}
\left(\frac{\partial^2}{\partial r^2}+\frac{k}{\pi \alpha^{\prime}}\right) R(r)=0
\label{reqn}
\end{align}
and
\begin{align}\nonumber
 \left[ \frac{16 \pi N}{\sin \theta} \frac{1}{\sqrt{ u v^2}}\left(u^2 v  \frac{\partial^2}{\partial u \partial v}-\frac{1}{2} u^3 \frac{\partial^2}{\partial u^2}-\frac{1}{2} u^2 \frac{\partial}{\partial u}-\frac{\sin^2\theta}{128\pi^2  \hbar^2}u^2v\right)\right.\\\left.- \pi \alpha^{\prime}\left(\frac{u}{ v}\frac{\partial^2}{\partial \theta^2}+\frac{u}{ v \sin^2\theta}\frac{\partial^2}{\partial \phi^2}\right)+\frac{ N^2 u }{4 \pi \alpha'} +k \right] \chi(u,v,\theta,\phi)=0,
  \label{uvtp}
\end{align}
where $-k$ is the separation constant.

Equation \ref{reqn} admits solutions oscillatory in $r$, given by
\begin{align}
 R(r)=c_1 e^{ i \sqrt{\frac{k}{\pi \alpha'}}r}+c_2 e^{- i \sqrt{\frac{k}{\pi \alpha'}}r},
\end{align}
where $c_1$ and $c_2$ are arbitrary constants.

In order to separate the $\phi$ coordinate, we first multiply equation \ref{uvtp} by $\frac{v}{u} \sin^2 \theta $, and write 
$ \chi (u,v,\theta,\phi)=\psi (u,v,\theta) \Phi (\phi)
$
which yields two differential equations of the forms
 \begin{align}
 \left(\frac{\partial^2}{\partial \phi^2}+\frac{\lambda}{\pi \alpha'}\right)\Phi(\phi)=0
 \label{phieqn}
 \end{align}
 and
 \begin{align}\nonumber
 \left[\frac{16 \pi N \sin \theta}{\sqrt{ u v^2}}\left(u v^2  \frac{\partial^2}{\partial u \partial v}-\frac{1}{2} u^2 v \frac{\partial^2}{\partial u^2}-\frac{1}{2} u v \frac{\partial}{\partial u}-\frac{\sin^2\theta}{128\pi^2 \hbar^2}uv^2\right)\right.\\\left.- \pi \alpha^{\prime}\sin^2 \theta \frac{\partial^2}{\partial \theta^2}+\frac{ N^2}{4 \pi \alpha'} v \sin^2 \theta+\frac{k v}{u} \sin^2 \theta  +\lambda\right]\psi(u,v,\theta)=0,
 \label{uvt}
 \end{align}
 where $-\lambda$ is the separation constant. 
 
Equation \ref{phieqn} admits solutions of the form 
\begin{align}
  \Phi(\phi)=c_3 e^{i\sqrt{\frac{\lambda}{\pi \alpha'}} \phi}+c_4 e^{-i\sqrt{\frac{\lambda}{\pi \alpha'}} \phi},
  \label{phi}
 \end{align}
 where $c_3$ and $c_4$ are arbitrary constants.
 This wave function must satisfy the periodicity condition $\Phi(\phi+2\pi)=\Phi(\phi)$, giving $\frac{\lambda}{\pi\alpha'}=m^2$, so that the separation constant $\lambda$ becomes quantized,
 \begin{align}
  \lambda_m=\pi\alpha' m^2,
  \label{lm}
 \end{align}
where $m=0,\pm1, \pm2, \ldots$.

Upon taking into account this quantization, $\lambda_m$ from equation \ref{lm}, the wave function given by \ref{phi} takes the form
\begin{align}
  \Phi_m(\phi)=c_3 e^{i m \phi}.
  \label{phim}
 \end{align}

 \subsection{Perturbation theory}

In equation \ref{uvt}, the variables $u$, $v$ and $\theta$ are intricately coupled, and hence it is not readily solvable. To simplify this equation, we employ the following transformation  
 \begin{align}
  \sqrt{uv}=z \quad {\rm and} \quad u=x,
  \label{tran}
 \end{align}
 whereby equation \ref{uvt} assumes the form  
  \begin{align}\nonumber
 \left[ \frac{8 \pi  N}{\sqrt{ x}}\left(\frac{z^2}{4} \frac{\partial^2}{\partial z^2}+\frac{z}{4} \frac{\partial}{\partial z}- x^2 \frac{\partial^2}{\partial x^2}- x  \frac{\partial}{\partial x}\right)-\pi \alpha^{\prime} \sin \theta \frac{\partial^2}{\partial \theta^2}  +\frac{\lambda}{\sin\theta}\right.\\\left.-  \frac{N}{8 \pi \hbar^2} \frac{z^2}{\sqrt{ x}} \sin^2\theta+\frac{ N^2}{4 \pi \alpha'} \frac{ z^2}{x} \sin \theta+\frac{k z^2}{x^2} \sin \theta  \right] \psi(x,z,\theta)=0.
 \label{zxt}
 \end{align}
 
 It is important to consider the behavior of this equation in the region near the classical singularity of the black hole. The classical singularity is determined by the classical expressions for the geometric variables, given by $u=\frac{2M}{t}-1$ and $v=t^2$, in the limit $t\to0$.  Consequently, the transformed variables behave like $z=\sqrt{uv}=\sqrt{2Mt-t^2}$ and $x=u=\frac{2M}{t}-1$. The last three terms in equation \ref{zxt} are composed of these variables. The nature of these terms near the singularity is controlled by the ratios $\frac{z^2}{\sqrt{x}}$, $\frac{z^2}{x}$ and $\frac{z^2}{x^2}$. These ratios have the behaviors $\frac{z^2}{\sqrt{x}}=\frac{2Mt^{3/2}-t^{5/2}}{\sqrt{2M-t}}\to0$, $\frac{z^2}{x}=\frac{2Mt^{2}-t^{3}}{2M-t}\to0$ and $\frac{z^2}{x^2}=\frac{2Mt^{3}-t^{4}}{(2M-t)^2}\to0$, near the singularity, where $t\to0$. Consequently, the last three terms in equation \ref{zxt} can be regarded as small in the region near the singularity. Thus, in order to find the behavior of the solution near the singularity, we can regard these terms as perturbation.

 To  set up the perturbation theory, we write equation \ref{zxt} in the form 
 \begin{align}
  (\hat H_0+\epsilon \hat H')\psi=0
  \label{pt}
 \end{align}
with the unperturbed Hamiltonian 
 \begin{align}
  \hat H_0= \frac{8 \pi  N}{\sqrt{ x}}\left(\frac{z^2}{4} \frac{\partial^2}{\partial z^2}+\frac{z}{4} \frac{\partial}{\partial z}- x^2 \frac{\partial^2}{\partial x^2}- x  \frac{\partial}{\partial x}\right)-\pi \alpha^{\prime} \sin \theta \frac{\partial^2}{\partial \theta^2}  +\frac{\lambda}{\sin\theta},
  \label{unpertH}
 \end{align}
and the perturbation 
 \begin{align}
   \hat H'=-  \frac{N}{8 \pi \hbar^2} \frac{z^2}{\sqrt{ x}} \sin^2\theta+\frac{ N^2}{4 \pi \alpha'}\frac{ z^2}{x} \sin \theta+\frac{k z^2}{x^2} \sin \theta,
  \label{pertH}
 \end{align}
 where $\epsilon$ is a book keeping parameter. 
 
 We shall solve equation  \ref{pt} perturbatively up to first order in $\epsilon$. Writing $\psi=\psi_0+\epsilon \psi_1+\ldots$, we have $(\hat H_0+\epsilon H')(\psi_0+\epsilon \psi_1+\ldots)=0$, so that
   \begin{align}
 \hat H_0\psi_0=0
  \label{pert0}
 \end{align}
 at $\mathcal{O}(\epsilon^0)$, and 
  \begin{align}
 \hat H_0 \psi_1 + H'\psi_0=0
  \label{pert1}
 \end{align}
  at $\mathcal{O}(\epsilon)$.

Importantly, since the perturbation Hamiltonian $H'$ approaches zero near the singularity, this perturbative framework provides a controlled approximation scheme for studying the deep quantum regime in the vicinity of the singularity.

 \section{Zeroth-order solution}
 \label{zero}
 
 Using the expression for the unperturbed Hamiltonian \ref{unpertH}, the zeroth order equation \ref{pert0} can be written explicitly as
\begin{align}
 \left[ \frac{8 \pi N}{\sqrt{ x}}\left(\frac{z^2}{4} \frac{\partial^2}{\partial z^2}+\frac{z}{4} \frac{\partial}{\partial z}- x^2 \frac{\partial^2}{\partial x^2}- x  \frac{\partial}{\partial x}\right)
 - \pi \alpha^{\prime} \sin \theta \frac{\partial^2}{\partial \theta^2} +\frac{\lambda}{\sin\theta}\right]\psi_0(x,z,\theta)=0,
 \label{0or}
\end{align}
where the angular part can be separated by writing
$
 \psi_0(x,z,\theta)=\tilde \psi_0(x,z) \Theta(\theta),
$
yielding two differential equations  
\begin{align}
 \left(\frac{z^2}{4} \frac{\partial^2}{\partial z^2}+\frac{z}{4} \frac{\partial}{\partial z}- x^2 \frac{\partial^2}{\partial x^2}- x  \frac{\partial}{\partial x}-\frac{\kappa}{8 \pi N}\sqrt{x}\right)\tilde \psi_0(x,z)=0
 \label{zx}
\end{align}
and 
\begin{align}
\left( \pi \alpha^{\prime} \sin \theta \frac{\partial^2}{\partial \theta^2} -\frac{\lambda}{\sin\theta}-\kappa\right)\Theta(\theta)=0,
\label{theta}
\end{align}
where $\kappa$ is the separation constant.

Equation \ref{zx} is further separable in $z$ and $x$ by writing 
$
 \tilde \psi_0(x,z)= X(x)Z(z),
$
 giving two differential equations,
\begin{align}
 \left( x^2 \frac{\partial^2}{\partial x^2}+ x  \frac{\partial}{\partial x}+\frac{\kappa}{8 \pi N}\sqrt{x}-2\gamma\right)X(x)=0
 \label{xeqn}
\end{align}
and
\begin{align}
 \left(z^2 \frac{\partial^2}{\partial z^2}+z \frac{\partial}{\partial z}-8\gamma \right)Z(z)=0,
 \label{zeqn}
\end{align}
where $\gamma$ is the separation constant. 

The zeroth order wave function is given by the solutions of \ref{xeqn}, \ref{zeqn} and \ref{theta}, so that
\begin{align}
 \psi_0(u,v,\theta)=X(x)Z(z)\Theta(\theta)=X(u)Z(\sqrt{uv})\Theta(\theta).
\end{align}

It is important to emphasize that the admissibility conditions for the wave function $\psi_0(u,v,\theta)$ must be carefully examined. One such condition is provided by the DeWitt criterion, which requires the wave function to vanish at the classical singularity. The classical minisuperspace variables are given by $
x=u=\frac{2M}{t}-1,$ and $z=\sqrt{uv}=\sqrt{2Mt-t^2},$ from which it follows that $x\to\infty$ and $z\to0$ as the singularity $t\to0$ is approached.

\subsection{The $Z(z)$ part}
\label{secz}

Equation \ref{zeqn} immediately admits solution of the form 
\begin{align}
Z(z)=  \left(c_5 z^{1+\sqrt{1+8 \gamma}}+c_6 z^{1-\sqrt{1+8 \gamma}}\right),
\label{Z}
\end{align}
 with $c_5$ and $c_6$ arbitrary constants.
 
 {\bf \underline{Case I: $\gamma>0$}}
 
 For $\gamma>0$, Eq. $\ref{Z}$ implies that the term $z^{,1-\sqrt{1+8\gamma}}$ diverges as $t\to0$. To ensure regularity of the wave function and satisfy the DeWitt criterion, we therefore set $c_6=0$, yielding
$
Z(z)=c_5 z^{1+\sqrt{1+8\gamma}},
$
which vanishes in the limit $t\to0$, satisfying the DeWitt criterion for singularity resolution.

For $\gamma>0$, the behavior of the wave function at the horizon $(t\to2M)$ must also be considered. Upon setting $c_6=0$ in equation \ref{Z}, we once again have $Z(z)\to0$ by virtue of $z\to0$ as $t\to2M$,  so that the overall wave function vanishes at the horizon. 

{\bf \underline{Case II: $\gamma<0$}}

For $\gamma<0$, the situation is qualitatively different. In this case, both contributions to $Z(z)$ vanish as the singularity is approached, $z\to0$, and therefore the DeWitt criterion is automatically satisfied, resolving the singularity. 

For $\gamma<0$,  $Z(z)\to0$ at the horizon $z\to0$, making the overall wave function vanish.

\subsection{The $X(x)$ part}
 
Equation \ref{xeqn} is the Bessel differential equation having the solution
\begin{align}
 X(x)=c_7 \Gamma (1+ \sqrt{32\gamma}) J_{ \sqrt{32\gamma }}\left(\textstyle \sqrt{\frac{2\kappa \sqrt{x}}{\pi N}}\right)
 +c_{8} \Gamma (1-\sqrt{32\gamma}) J_{ -\sqrt{32\gamma }}\left(\textstyle \sqrt{\frac{2\kappa \sqrt{x}}{\pi N}}\right),
 \label{X}
\end{align}
where $J_{\nu}(\xi)$ is the Bessel function of the first kind of order $\nu=\pm \sqrt{32\gamma }$,  while $c_7$ and $c_{8}$ are arbitrary constants.

{\bf \underline{Case I: $\gamma>0$}}

In the regime $\gamma>0$, Eq. $\ref{X}$ shows that the Bessel functions $J_{\pm\sqrt{32\gamma}}(\beta x^{1/4})$ vanish as $x\to\infty$, corresponding to the classical singularity \(t\to0\), reinforcing   singularity resolution.

Moreover, for $\gamma>0$,   the Bessel function $J_{-\sqrt{32\gamma}}(\beta x^{1/4})$ is divergent (or ill-defined) for non-integer values of $\sqrt{32\gamma}$, at the horizon where $x\to0$, whereas it remains finite only when $\sqrt{32\gamma}$ is an integer. Consequently, regularity of the wave function at the horizon requires a discrete spectrum for $\gamma$, namely
\begin{align}
\gamma_n=\frac{n^2}{32},
\label{gn}
\end{align}
where $n\in\mathbb{Z}$.

Upon taking into account the quantization of the separation constant $\gamma_n$ from equation \ref{gn}, the wave functions given by \ref{X}  becomes 
\begin{align}
 X_n(x)=c_7 \Gamma (1+ n) J_{ n }\left(\textstyle \sqrt{\frac{2\kappa \sqrt{x}}{\pi N}}\right)+c_{8} \Gamma (1-n) J_{ -n}\left(\textstyle \sqrt{\frac{2\kappa \sqrt{x}}{\pi N}}\right).
 \label{Xn}
\end{align}
Since $n \in \mathbb{Z}$, the gamma function $\Gamma(1-n)$ is singular for $n>0$. To ensure that the solution remains well defined, we must therefore set $c_8=0$ in this case. Likewise, $\Gamma(1+n)$ becomes singular for $n<0$, requiring $c_7=0$.

Moreover, the quantization of the separation constant $\gamma_n$ in equation \ref{gn} gives the $Z(z)$ part of the wave function \ref{Z} in the form
\begin{align}
Z_n(z)=  \left(c_5 z^{1+\sqrt{1+\frac{1}{4}n^2}}+c_6 z^{1-\sqrt{1+\frac{1}{4}n^2}}\right).
\label{Zn}
\end{align}

{\bf \underline{Case II: $\gamma<0$}}

 In the regime $\gamma<0$, the Bessel functions appearing in $X(x)$ \ref{X} acquire imaginary order, and they approach zero at the singularity, $x\to\infty$, further ensuring  singularity resolution. 

  However, at the horizon $x\to0$, both Bessel functions in $X(x)$ remain finite and generally take complex values.

  \subsection{The $\Theta(\theta)$ part}

Equation \ref{theta} is the Heun differential equation yielding the solution
\begin{align}\nonumber
 &\Theta(\theta)=c_{9}\ \text{HeunG}\left(2,\textstyle \frac{\kappa}{\pi  \alpha' }-\frac{\lambda }{\pi  \alpha' },-i\sqrt{\frac{\kappa}{\pi\alpha'}},i\sqrt{\frac{\kappa}{\pi\alpha'}},\frac{1}{2},0,s\right)\\+&c_{10}\  \sqrt{s} \  \text{HeunG}\left(2,\textstyle \frac{\kappa}{\pi  \alpha' }-\frac{\lambda }{\pi  \alpha' }+\frac{1}{4},\frac{1}{2}+i\sqrt{\frac{\kappa}{\pi\alpha'}},\frac{1}{2}-i\sqrt{\frac{\kappa}{\pi\alpha'}},\frac{3}{2},0,s\right),
 \label{T}
\end{align}
where $\text{HeunG}(k_1,k_2,k_3,k_4,k_5,k_6,s)$ is the general Heun function with $s=1+\sin\theta$,  while $c_{9}$ and $c_{10}$ are arbitrary constants.

  \subsection{Zeroth-order wave function}

Upon employing the functional forms given by equations \ref{X},  \ref{Z} and \ref{T}, we obtain the zeroth order wave function as
 \begin{align}\nonumber
 &\psi_{0}(z,x,\theta)= \left(c_5 z^{a_+}+c_6 z^{a_-}\right)\left\{c_7 \Gamma (b_+) J_{\nu}\left(\textstyle \beta x^{\frac{1}{4}}\right)+c_{8} \Gamma (b_-) J_{ -\nu}\left(\textstyle \beta x^{\frac{1}{4}}\right)\right\}\\\nonumber&\left\{c_{9}\ \text{HeunG}\left(2,\textstyle \frac{\kappa}{\pi  \alpha' }-\frac{\lambda }{\pi  \alpha' },-i\sqrt{\frac{\kappa}{\pi\alpha'}},i\sqrt{\frac{\kappa}{\pi\alpha'}},\frac{1}{2},0,s\right)\right.\\
 &\left.+c_{10}\sqrt{s} \  \text{HeunG}\left(2,\textstyle \frac{\kappa}{\pi  \alpha' }-\frac{\lambda }{\pi  \alpha' }+\frac{1}{4},\frac{1}{2}+i\sqrt{\frac{\kappa}{\pi\alpha'}},\frac{1}{2}-i\sqrt{\frac{\kappa}{\pi\alpha'}},\frac{3}{2},0,s\right)\right\},
 \label{psi0}
 \end{align}
which assumes different forms for the cases $\gamma>0$ and $\gamma<0$.
Here $a_{\pm}=1\pm\sqrt{1+8\lambda}$, $b_{\pm}=(1\pm\sqrt{32\gamma})$, $\beta=\sqrt{\frac{2\kappa}{\pi N}}$  and $\nu=\sqrt{32\gamma}$.

As described in Sec.~\ref{secz}, this wavefunction vanishes at the singularity (with appropriate choice of constants), which is the DeWitt boundary condition for singularity resolution. Moreover, the wave function at the horizon as well.

\section{First-order correction}
\label{FOC}

 The first order equation \ref{pert1}, upon using the expressions for $\hat H_0$ and $\hat H'$ from equations \ref{unpertH} and \ref{pertH}, assumes the explicit form 
\begin{align}\nonumber
\left[\frac{8 \pi  N}{\sqrt{ x}}\left(\frac{z^2}{4} \frac{\partial^2}{\partial z^2}+\frac{z}{4} \frac{\partial}{\partial z}- x^2 \frac{\partial^2}{\partial x^2}- x  \frac{\partial}{\partial x}\right)-\pi \alpha^{\prime} \sin \theta \frac{\partial^2}{\partial \theta^2}  +\frac{\lambda}{\sin\theta}\right]\psi_1\\=\left[\frac{N}{8 \pi \hbar^2} \frac{z^2}{\sqrt{ x}} \sin^2\theta-\frac{ N^2}{4 \pi \alpha'}\frac{ z^2}{x} \sin \theta-\frac{k z^2}{x^2} \sin \theta\right]\psi_0,
 \label{1or}
\end{align}
where the right-hand side acts as a source term for $\psi_1$.

\subsection{Complementary function}

Since the operator on the left-hand side of equation \ref{1or} has the same form as that of \ref{0or}, the complementary function for $\psi_1$ will have the same form as the solution for $\psi_0$. Consequently, the complementary function for equation \ref{1or} can be written immediately,  following \ref{psi0}, as   
\begin{align}\nonumber
 &\psi_{1}^{\rm CF}= \left(c'_5 z^{a_+'}+c'_6 z^{a_-'}\right)\left\{c'_7 \Gamma (b_+') J_{ \nu' }\left(\textstyle \beta'x^{\frac{1}{4}}\right)+c'_{8} \Gamma (b_-') J_{-\nu}\left(\textstyle \beta'x^{\frac{1}{4}}\right)\right\}\\\nonumber&\left\{c'_{9}\ \text{HeunG}\left(2,\textstyle \frac{\kappa'}{\pi  \alpha' }-\frac{\lambda }{\pi  \alpha' },-i\sqrt{\frac{\kappa'}{\pi\alpha' }},i\sqrt{\frac{\kappa'}{\pi\alpha' }},\frac{1}{2},0,s\right)\right.\\&\left.+c'_{10}\  \sqrt{s} \  \text{HeunG}\left(2,\textstyle \frac{\kappa'}{\pi  \alpha' }-\frac{\lambda }{\pi  \alpha' }+\frac{1}{4},\frac{1}{2}+i\sqrt{\frac{\kappa'}{\pi\alpha' }},\frac{1}{2}-i\sqrt{\frac{\kappa'}{\pi\alpha' }},\frac{3}{2},0,s\right)\right\},
 \label{cf}
 \end{align}
 where $\gamma'$ and $\kappa'$ are separation constants and $c'_i$ $(i=5,6,\ldots,10)$ are arbitrary constants.

 {\bf \underline{Case I: $\gamma'>0$}}
 
 Similarly to the zeroth order wave function, 
 for $\gamma'>0$, Eq. $\ref{cf}$ implies that the term $z^{1-\sqrt{1+8\gamma'}}$ diverges as $t\to0$. To ensure regularity of the wave function and satisfy the DeWitt criterion, we therefore set $c'_6=0$, yielding the first factor of the complementary function as 
$
c'_5 z^{1+\sqrt{1+8\gamma'}},
$
which vanishes in the limit $t\to0$, satisfying the DeWitt criterion for singularity resolution.

Moreover, in the regime $\gamma'>0$, Eq. \ref{cf} shows that the Bessel functions $J_{\pm\sqrt{32\gamma'}}(\beta' x^{1/4})$ vanish as $x\to\infty$, corresponding to the classical singularity \(t\to0\), reinforcing   singularity resolution.

For $\gamma'>0$, the behavior of the wave function at the horizon $(t\to2M)$ must also be considered. Upon setting $c'_6=0$ in equation \ref{cf}, we once again have a vanishing $z$-part by virtue of $z\to0$ as $t\to2M$,  so that the overall complementary function vanishes at the horizon.

In addition, for $\gamma'>0$,   the Bessel function $J_{-\sqrt{32\gamma'}}(\beta' x^{1/4})$ is divergent (or ill-defined) for non-integer values of $\sqrt{32\gamma'}$, at the horizon where $x\to0$, whereas it remains finite only when $\sqrt{32\gamma'}$ is an integer. Consequently, regularity of the wave function at the horizon requires a discrete spectrum for $\gamma'$, namely
\begin{align}
\gamma'_{n'}=\frac{n'^2}{32},
\label{gn2}
\end{align}
where $n'\in\mathbb{Z}$.

{\bf \underline{Case II: $\gamma'<0$}}

For $\gamma'<0$, the situation is qualitatively different. In this case, both terms in the $z$-part of the complementary function \ref{cf} vanish as the singularity is approached, $z\to0$, and therefore the DeWitt criterion is automatically satisfied, resolving the singularity.

 In the regime $\gamma'<0$, the Bessel functions appearing in the $x$-part of the complementary function \ref{cf} acquire imaginary order, and they approach zero at the singularity, $x\to\infty$, further ensuring  singularity resolution.

 At the horizon $z\to0$,  and for $\gamma'<0$, both terms in the $z$-part vanish at the horizon $z\to0$, making the complementary function \ref{cf} vanish.

  However, at the horizon $x\to0$, both Bessel functions in the complementary function \ref{cf} remain finite and generally take complex values.

 {\bf \underline{Case I: $\gamma'>0$}}

Analogous to the zeroth-order wave function, when $\gamma' > 0$, Eq. (\ref{cf}) indicates that the term $z^{1-\sqrt{1+8\gamma'}}$ diverges as $t \to 0$. To maintain the regularity of the wave function and satisfy the DeWitt criterion, we therefore set $c'_6 = 0$. This yields the first factor of the complementary function:
$
c'_5 z^{1+\sqrt{1+8\gamma'}},
$
which tends to zero in the limit $t \to 0$, ensuring compliance with the DeWitt condition for singularity resolution.

Furthermore, for $\gamma' > 0$, Eq. (\ref{cf}) demonstrates that the Bessel functions $J_{\pm\sqrt{32\gamma'}}(\beta' x^{1/4})$ vanish as $x \to \infty$, corresponding to the classical singularity at $t \to 0$, thus reinforcing the resolution of the singularity.

The  behavior of the complementary function near the horizon $(t \to 2M)$ also requires attention. With $c'_6 = 0$, the $z$-dependent part vanishes as $z \to 0$ when $t \to 2M$, causing the overall complementary function to vanish at the horizon as well.

Additionally, the Bessel function $J_{-\sqrt{32\gamma'}}(\beta' x^{1/4})$ diverges or becomes ill-defined for non-integer values of $\sqrt{32\gamma'}$ at the horizon $x \to 0$. Regularity at the horizon therefore imposes a quantization condition on $\gamma'$,
\begin{align}
\gamma'_{n'} = \frac{n'^2}{32}, \quad n' \in \mathbb{Z}.
\label{gn2}
\end{align}

{\bf \underline{Case II: $\gamma'<0$}}

For $\gamma' < 0$, the behavior of the system changes fundamentally. In this regime, both terms in the $z$-dependent part of the complementary function (\ref{cf}) vanish as $z \to 0$ near the singularity, automatically fulfilling the DeWitt criterion and thereby resolving the singularity.

In addition, the Bessel functions in the $x$-dependent part acquire imaginary orders and approach zero as $x \to \infty$, further guaranteeing singularity resolution.

At the horizon, $z \to 0$, both $z$-dependent terms vanish for $\gamma' < 0$, so the complementary function again tends to zero.

However, when $x \to 0$ at the horizon, the Bessel functions in the complementary function remain finite and typically assume complex values.

\subsection{Particular integral}

The particular integral of equation \ref{1or} requires special consideration as the source term on the right-hand side contains the zeroth order solution $\psi_0$, which is a product of special functions: Bessel and general Heun function. 

Multiplying equation \ref{1or} throughout by $\sqrt{x}$ and making the transformations 
\begin{align}
 w=\ln x \quad {\rm and} \quad y=\ln z,
 \label{wy}
\end{align}
we obtain  
\begin{align}\nonumber
 \left[8\pi N \left(\frac{1}{4} \frac{\partial^2}{\partial y^2}-  \frac{\partial^2}{\partial w^2}\right)- e^{w/2} \left(\pi \alpha^{\prime} \sin \theta \frac{\partial^2}{\partial \theta^2} - \frac{\lambda}{\sin\theta}\right)\right]\psi_1\\=e^{2y} \sin\theta \left[  \frac{N}{8\pi \hbar^2} \sin\theta -\frac{ N^2}{4 \pi \alpha'} e^{-w/2} -k  e^{-3w/2}\right]\psi_0.
 \label{1or2}
\end{align}

Since it is important to consider the behavior near the classical singularity, we assess the magnitude of the terms in equation \ref{1or2} from the classical expressions $u=\frac{2M}{t}-1$ and $v=t^2$  near the singularity, $t\to0$. Thus, $w=\ln x=\ln u=\ln (\frac{2M}{t}-1)\to \infty$ and $y=\ln z=\frac{1}{2}\ln (uv)=\frac{1}{2}\ln (2Mt-t^2)\to-\infty$, at the singularity, $t\to0$. Consequently, the right-hand side of equation \ref{1or2} vanishes in this limit, whereas the second term on the left-hand side approaches infinity near the singularity. Hence, the second term on the left-hand side gives the most dominant contribution and we write an auxiliary equation 
\begin{align}
 - e^{w/2} \left(\pi \alpha^{\prime} \sin \theta \frac{\partial^2}{\partial \theta^2} - \frac{\lambda}{\sin\theta}\right)f(\theta)=0.
 \label{1or0}
\end{align}

Equation \ref{1or0} is the Legendre differential equation whose solution is given  by 
\begin{align}
 f(\theta)=\left\{c_{11}  \, P_{-\frac{1}{2}}^{\sqrt{\frac{1}{4} +\frac{ \lambda }{\pi\alpha' }}}(\cos \theta)+c_{12}\,    Q_{-\frac{1}{2}}^{\sqrt{\frac{1}{4} +\frac{ \lambda }{\pi\alpha' }}}(\cos \theta)\right\}\sqrt{\sin\theta },
\end{align}
where $P_{\nu}^{\mu}(\xi)$ and $Q_{\nu}^{\mu}(\xi)$ are the associated Legendre functions of the first and second kind of degree $\nu=-\frac{1}{2}$ and order $\mu=\sqrt{\frac{1}{4} +\frac{ \lambda }{\pi\alpha' }}$, with  $c_{11}$ and $c_{12}$ arbitrary constants.

Now, expressing the first order solution as $\psi_1(y,w,\theta)=F(y,w)f(\theta)$, and substituting in equation \ref{1or2}, and further using \ref{1or0}, we obtain 
\begin{align}
 8\pi N f(\theta){\textstyle \left(\frac{1}{4} D^2_y-  D^2_w\right)} F_{\theta}(y,w)
 =e^{2y} \sin\theta \left[  \frac{N \sin\theta}{8\pi \hbar^2}  -\frac{ N^2}{4 \pi \alpha'} e^{-w/2} -k  e^{-3w/2}\right]\psi_0(y,w,\theta),
 \label{1or3}
\end{align}
where the differential operators are defined as $D_y \equiv \frac{d}{dy} $ and $ D_w \equiv \frac{d}{dw}$.

Since there is no differential operator with respect to $\theta$ in equation \ref{1or3}, $\theta$ appears as a parameter (hence the notation $F_{\theta}$) in  this equation. Moreover, the functional forms of $\psi_0(y,w,\theta)$ and $f(\theta)$ are known from the preceding calculations, and the function $F_{\theta}(y,w)$ is required to be determined from the differential equation \ref{1or3}.

The particular integral of equation \ref{1or2} is therefore given by
\begin{align}
 \psi_1^{\rm PI}(y,w,\theta)=f(\theta) F_{\theta}^{\rm PI}(y,w),
 \label{psi1PI}
\end{align}
where $ F_{\theta}^{\rm PI}(y,w)$ is the particular integral of equation \ref{1or3} that can be calculated from 
\begin{align}
 F^{\rm PI}_{\theta}(y,w)= \frac{1}{8\pi N }\frac{ \sin\theta}{f(\theta)} {\textstyle \left(\frac{1}{4} D^2_y-  D^2_w\right)^{-1}} e^{2y}
    \left[  \frac{N  \sin\theta }{8\pi \hbar^2}-\frac{ N^2}{4 \pi \alpha'} e^{-w/2} -k  e^{-3w/2}\right]\psi_0(y,w,\theta),
 \label{1or4}
\end{align}

Upon using $\psi_0(w,y,\theta)=\tilde \psi_0(w,y) \Theta(\theta)=\tilde \psi_0(x,z)\Theta(\theta)=X(x)Z(z)\Theta(\theta)$ from the solutions \ref{Z}, \ref{X} and \ref{T} in equation \ref{1or4}, we have  
\begin{align}
 F^{\rm PI}_{\theta}(y,w)= \frac{1}{8\pi N }\frac{\Theta(\theta) \sin\theta}{f(\theta)}  \left[  \frac{N \sin\theta}{8\pi \hbar^2}  I_0 -\frac{ N^2}{4 \pi \alpha'} I_1 -k  I_3 \right],
 \label{PIF}
\end{align}
where 
\begin{align}\nonumber
 I_0&= {\textstyle \left(\frac{1}{4} D^2_y-  D^2_w\right)^{-1}} e^{2y}\tilde \psi_0(y,w), \\ \label{Ins} I_1&={\textstyle \left(\frac{1}{4} D^2_y-  D^2_w\right)^{-1}} e^{2y} e^{-w/2}\tilde \psi_0(y,w), 
 \\\nonumber  I_3&={\textstyle \left(\frac{1}{4} D^2_y-  D^2_w\right)^{-1}} e^{2y} e^{-3w/2}\tilde \psi_0(y,w),
\end{align}
that can be calculated explicitly.

\subsection{Calculation of $I_j$}

To evaluate the three terms in equations \ref{Ins},  we compute the general integral  
\begin{align}
I_j= {\textstyle \left(\frac{1}{4} D^2_y-  D^2_w\right)^{-1}} e^{2y} e^{-jw/2}\tilde \psi_0(y,w).
\end{align}
 Thus, upon substituting from \ref{Z} and \ref{X}, with \ref{wy}, we have 
 \begin{align}\nonumber
 &I_j={\textstyle\left(\frac{1}{4} D^2_y -  D^2_w \right)^{-1}} e^{-jw/2}  \left[C_1 e^{\left(3-\frac{1}{2}\sqrt{\nu^2+4}\right)y} \Gamma \left(\nu+1\right)   J_{\nu}(\beta e^{w/4})\right.\\&\left.+C_2 e^{\left(3+\frac{1}{2}\sqrt{\nu^2+4}\right)y} \Gamma \left(\nu+1\right)  J_{\nu}(\beta e^{w/4})+C_3 e^{\left(3-\frac{1}{2}\sqrt{\nu^2+4}\right)y}  \Gamma \left(1-\nu\right) J_{-\nu}(\beta e^{w/4})\right.\\&\left. +C_4 e^{\left(3+\frac{1}{2}\sqrt{\nu^2+4}\right)y} \Gamma \left(1-\nu\right) J_{-\nu}(\beta e^{w/4}) \right],
 \label{Inc}
\end{align}
recalling that $\nu=\sqrt{32 \gamma}$, and we have denoted $\sqrt{\frac{2\kappa}{\pi N}}=\beta$, and $C_1=c_6c_7$, $C_2=c_5c_7$, $C_3=c_6c_{8}$ and $C_4=c_5c_{8}$.

Noting that the Bessel function admits the expansion
$
 J_{\nu}(x)=\sum_{l=0}^{\infty} \frac{(-1)^l}{l! \Gamma(l+\nu+1)}\left(\frac{x}{2}\right)^{2l+\nu}
$, the first, second, third, and the fourth terms in equation \ref{Inc} take the forms
\begin{align}\nonumber
& {\textstyle\left(\frac{1}{4} D^2_y -  D^2_w \right)^{-1}}  e^{\left(3-\frac{1}{2}\sqrt{\nu^2+4}\right)y} e^{-jw/2} J_{\nu}(\beta e^{w/4})=e^{\left(3-\frac{1}{2}\sqrt{\nu^2+4}\right)y}  \\&\times\sum_{l=0}^{\infty} \frac{e^{w (\frac{l}{2}+\frac{\nu}{4}-\frac{j}{2})}}{\left[\frac{1}{4} \left(3-\frac{1}{2}\sqrt{\nu^2+4}\right)^2 - \left(\frac{l}{2}+\frac{\nu}{4}-\frac{j}{2}\right)^2 \right]}\frac{(-1)^l}{l! \Gamma(l+\nu+1)}\left(\frac{\beta }{2}\right)^{2l+\nu},
  \label{i21}
\end{align}
\begin{align}\nonumber
 &{\textstyle\left(\frac{1}{4} D^2_y -  D^2_w \right)^{-1}}  e^{\left(3+\frac{1}{2}\sqrt{\nu^2+4}\right)y} e^{-jw/2} J_{\nu}(\beta e^{w/4})=e^{\left(3+\frac{1}{2}\sqrt{\nu^2+4}\right)y} \\&\times \sum_{l=0}^{\infty} \frac{e^{w (\frac{l}{2}+\frac{\nu}{4}-\frac{j}{2})}}{\left[\frac{1}{4} \left(3+\frac{1}{2}\sqrt{\nu^2+4}\right)^2 -  \left(\frac{l}{2}+\frac{\nu}{4}-\frac{j}{2}\right)^2 \right]}\frac{(-1)^l}{l! \Gamma(l+\nu+1)}\left(\frac{\beta }{2}\right)^{2l+\nu},
  \label{i22}
\end{align}
\begin{align}\nonumber
&{\textstyle\left(\frac{1}{4} D^2_y -  D^2_w \right)^{-1}}  e^{\left(3-\frac{1}{2}\sqrt{\nu^2+4}\right)y}  e^{-jw/2} J_{-\nu}(\beta e^{w/4})=e^{\left(3-\frac{1}{2}\sqrt{\nu^2+4}\right)y} \\& \times \sum_{l=0}^{\infty} \frac{e^{w (\frac{l}{2}-\frac{\nu}{4}-\frac{j}{2})}}{\left[\frac{1}{4} \left(3-\frac{1}{2}\sqrt{\nu^2+4}\right)^2 -  \left(\frac{l}{2}-\frac{\nu}{4}-\frac{j}{2}\right)^2 \right]}\frac{(-1)^l}{l! \Gamma(l-\nu+1)}\left(\frac{\beta }{2}\right)^{2l-\nu},
 \label{i23}
\end{align}
and
\begin{align}\nonumber
&{\textstyle\left(\frac{1}{4} D^2_y -  D^2_w \right)^{-1}} e^{\left(3+\frac{1}{2}\sqrt{\nu^2+4}\right)y} e^{-jw/2} J_{-\nu}(\beta e^{w/4})=e^{\left(3+\frac{1}{2}\sqrt{\nu^2+4}\right)y} \\&\times \sum_{l=0}^{\infty} \frac{e^{w (\frac{l}{2}-\frac{\nu}{4}-\frac{j}{2})}}{\left[\frac{1}{4} \left(3+\frac{1}{2}\sqrt{\nu^2+4}\right)^2 -  \left(\frac{l}{2}-\frac{\nu}{4}-\frac{j}{2}\right)^2 \right]}\frac{(-1)^l}{l! \Gamma(l-\nu+1)}\left(\frac{\beta }{2}\right)^{2l-\nu}.
 \label{i24}
\end{align}

As follows from the particular-integral contributions in Eqs.~\eqref{i21}-\eqref{i24}, the denominators can vanish for specific values of the summation index $l$ for the case $\nu=0$. Thus, the particular integral $F^{\rm PI}_{\theta}(y,w)$, which contributes to the wave function $\psi_1^{\rm PI}(y,w,\theta)$, would diverge if $\nu=0$, due to a vanishing denominator in the series.  Consequently, we restrict the parameter $\nu$ to non-zero values, $\nu\neq0$, to obtain physical solutions.

We now examine the behaviour of the particular integral (\ref{PIF}) in the vicinity of the classical singularity, $t\to0$, using the classical relations $u=\frac{2M}{t}-1$ and $v=t^2$. In this limit,
$
w=\ln x=\ln u=\ln\left(\frac{2M}{t}-1\right)\to\infty,
$
while
$
y=\ln z=\frac{1}{2}\ln(uv)
=\frac{1}{2}\ln(2Mt-t^2)\to-\infty.
$
Consequently, the arguments of all four Bessel functions appearing in the particular integral become asymptotically large, and hence
$
J_{\pm\nu}\left(\beta e^{w/4}\right)\to0,
$
as $t\to0$. It follows that all $I_j$ in Eq. (\ref{Inc}) vanish in this limit, implying
$
F^{\rm PI}_{\theta}(y,w)\to0.
$
Therefore, the particular-integral contribution to the first-order wave function,
$
\psi_1^{\rm PI}(y,w,\theta),
$
also vanishes at the singularity. Hence, the DeWitt boundary condition is satisfied by the particular integral.

We next investigate the behaviour of the particular integral near the horizon, $t\to2M$. Employing again the classical relations $u=\frac{2M}{t}-1$ and $v=t^2$, one finds
$
w=\ln x=\ln u
=\ln\left(\frac{2M}{t}-1\right)\to-\infty,
$
and
$
y=\ln z=\frac{1}{2}\ln(uv)
=\frac{1}{2}\ln(2Mt-t^2)\to-\infty.
$

In this limit, the argument of the Bessel functions tends to zero. For $\nu>0$, the function
$
J_{\nu}\left(\beta e^{w/4}\right)
$
vanishes at the horizon. By contrast,
$
J_{-\nu}\left(\beta e^{w/4}\right)
$
remains regular only when $\nu$ is an integer; otherwise it becomes singular. Regularity of the wave function at the horizon therefore requires $\nu\in\mathbb{Z}$, which in turn leads to the quantization condition on the parameter $\gamma$.

Under this condition, all terms entering the particular integral remain finite, ensuring that the first-order particular-integral contribution,
$
\psi_1^{\rm PI}(y,w,\theta),
$
is well defined and regular at the horizon.

\subsection{Total wave function}

Combining all the above results near the singularity, namely, zeroth-order, and complementary-function and particular-integral contributions in the first-order, we find that the total wave function,
\begin{equation}
\psi=\psi_0+\psi_1
=\psi_0+\psi_1^{\rm CF}+\psi_1^{\rm PI},
\end{equation}
approaches zero as the classical singularity is approached, $t\to0$. The DeWitt criterion is therefore fulfilled, indicating that the singularity is resolved, up to first order in the perturbative expansion.

\section{Discussion and conclusion}
\label{DC}

In this work, we investigated the quantum dynamics of the Schwarzschild black hole interior geometry  coupled with a string propagating in the interior. Starting from the Einstein-Hilbert action for the gravitational sector described by the Kantowski-Sachs metric and the Polyakov action for the string sector, we constructed a Hamiltonian formulation of the coupled geometry-string system.

Subsequently, we quantized this geometry-string system in the framework of canonical quantum gravity. The resulting Wheeler-DeWitt equation naturally incorporates  the black hole minisuperspace degrees of freedom coupled with the dynamical degrees of freedom of the string. This approach provided a unified description in which geometric and string degrees of freedom are quantized simultaneously.

The motivation for incorporating string degrees of freedom stems from the expectation that the region in the vicinity of the classical singularity corresponds to a regime of extremely high spacetime curvature, where conventional point-particle descriptions should break down and string-theoretic dgrees of freedom are expected to be awakened. In this deep quantum regime, the string degrees of freedom can effectively probe the Planckian or string scales so that the quantum nature of the region at the otherwise classical singularity can be assessed appropriately.

The Wheeler-DeWitt equation is found to be separable in the radial and azimuthal string sectors, with oscillatory solutions for both, the latter leading to a quantization condition on one of the separation constants due to periodic boundary condition in the azimuthal direction.

The remaining geometric and string degrees of freedom continue to be nontrivially coupled. To address this challenge, we developed a perturbative treatment. Reexpressing in suitable variables and by examining the behaviour near the singularity, we found that several terms in the Hamiltonian approach zero near the singularity and thus may be treated perturbatively. This observation provides a controlled approximation scheme for studying the deep quantum regime in the vicinity of the singularity.

 The zeroth-order wave function exhibits several important properties. The angular string sector is governed by the general Heun function, while the geometric Kantowski-Sachs sector is described by the Bessel function. Analysis of these solutions demonstrates that the wave function vanishes as the classical singularity is approached. Consequently, the DeWitt boundary condition is satisfied, indicating that the singular configuration is non-existent in the physical quantum state space. This result suggests that the black hole singularity is resolved at the quantum level of the zeroth order wave function.

 The first-order perturbative analysis provides an important consistency check of the singularity-resolution mechanism. The perturbation originates from terms that become negligible near the classical singularity but nevertheless contribute correction to the zeroth order wave function. In this perturbative analysis, the first-order correction is determined by an inhomogeneous equation whose source term turns out to contain the zeroth-order solution. Solving the inhomogeneous equation revealed that the complementary-function contribution obeys the same qualitative behaviour as the zeroth-order solution.  Consequently, the complementary function also satisfies the DeWitt criterion for singularity resolution.
 
 In addition to an explicit expression for the complementary function,  the solution also yields an explicit expression for the particular integral, providing a complete characterization of the first-order quantum state. Moreover, the particular integral also vanishes in the singular limit because the associated Bessel functions decay asymptotically near the singularity. As a result, the full wave function, including first-order corrections, continues to satisfy the DeWitt criterion. The persistence of this behaviour beyond leading order indicates that singularity resolution is not merely restricted to the zeroth-order wave function.

An important outcome of our analysis is the emergence of quantization conditions from the requirement of regularity of the wave function.  Demanding that the wave function remains finite at the horizon imposes an additional quantization condition on an eigenvalue occurring as the order of the Bessel function. This indicates that the quantum structure of the black hole interior naturally gives rise to a discrete spectrum, reflecting the restrictions imposed by the underlying geometry and boundary conditions. Thus, the quantum theory simultaneously constrains the behaviour of the wave function at two physically significant regions of the spacetime: the classical singularity and the event horizon. Such a dual role of boundary conditions provides  insights into the structure of physically acceptable quantum states inside black holes.

The central result of the present work concerns the fate of the classical singularity. We have shown that the zeroth-order wave function vanishes in the singular limit and therefore satisfies the DeWitt boundary condition. More significantly, the first-order correction, including both its complementary function and particular-integral contributions, exhibits the same behaviour. Consequently, the complete wave function up to first order approaches zero at the classical singularity. Within this perturbative framework, this behaviour implies resolution of the black hole singularity in the quantum theory.

In summary, we have presented a quantum gravity description of the black hole interior with the string degrees of freedom in order to probe the deep quantum regime near the singularity. We obtained explicit perturbative solutions to the corresponding Wheeler-DeWitt equation up to first order. The resulting wave functions remain regular at the horizon and satisfy the DeWitt boundary condition at the classical singularity. Overall, these findings indicate that the incorporation of string degrees of freedom into a canonical quantum-gravitational framework provides a viable mechanism for resolving the classical singularity in the black hole interior. The persistence of the DeWitt condition beyond leading order strengthens this conclusion.

\section*{Acknowledgments}
Harpreet Singh is supported by the Prime Minister’s Research Fellowship (PMRF ID: 1902180), Ministry of Education, Government of India.


\end{document}